\documentclass[journal]{IEEEtran}
\usepackage{amsmath,amsfonts}
\usepackage{array}
\usepackage{textcomp}
\usepackage{stfloats}
\usepackage{xurl} 
\usepackage{verbatim}
\usepackage{cite}
\usepackage{mathrsfs}
\usepackage{hyperref}
\usepackage{siunitx}

\usepackage{graphicx}
\usepackage{float}
\usepackage{algorithm}
\usepackage{algpseudocode}
\usepackage{tabularx}
\usepackage{array}
\newcolumntype{L}{>{\raggedright\arraybackslash}X}
\usepackage{booktabs}
\usepackage[table]{xcolor}
\usepackage{etoolbox}
\usepackage{xstring}
\usepackage[intoc]{nomencl}
\renewcommand{\nomgroup}[1]{%
  \item[\bfseries
    \ifstrequal{#1}{S}{Sets}{%
    \ifstrequal{#1}{P}{Parameters}{%
    \ifstrequal{#1}{V}{Variables}{Other}}}]%
}

\makenomenclature
\nomenclature[S]{$\mathscr{T}$}{Time steps of the planning horizon.}
\nomenclature[S]{$\mathscr{N}$}{Network buses.}
\nomenclature[S]{$\mathscr{L}$}{Network branches.}
\nomenclature[S]{$\mathscr{G}$}{Power generators.}
\nomenclature[S]{$\mathscr{S}$}{BESSs.}
\nomenclature[V]{{\footnotesize$\operatorname{CAPEX}$}}{Capital expenditure.}
\nomenclature[V]{{\footnotesize$\operatorname{OPEX}_d$}}{Operational expenditure for subproblem $d$.}
\nomenclature[V]{$U_{s}$}{Site variables of BESS $s$.}
\nomenclature[V]{$C_{s}, W_{s}$}{Size (energy and power) variables of BESS $s$.}
\nomenclature[V]{$P_{n_s,t}$}{Active power of BESS $s$ at node $n$, time $t$.}
\nomenclature[V]{$Q_{n_s,t}$}{Reactive power of BESS $s$ at node $n$, time $t$.}
\nomenclature[V]{$E_{s,t}$}{State of energy of BESS $s$ at node $n$, time $t$.}
\nomenclature[V]{$P_{n_g,t}$}{Active power of generator $g$ at node $n$, time $t$.}
\nomenclature[V]{$Q_{n_g,t}$}{Reactive power of generator $g$ at node $n$, time $t$.}
\nomenclature[V]{$V_{n,t}$}{Voltage magnitude at node $n$, time $t$.}
\nomenclature[V]{$\theta_{n,t}$}{Voltage phase at node $n$, time $t$.}
\nomenclature[V]{$p_{n,t}$}{Active power at node $n$, time $t$.}
\nomenclature[V]{$q_{n,t}$}{Reactive power at node $n$, time $t$.}
\nomenclature[V]{$p_{s_l,t}$}{Active power at sending end $s$ of branch $l$, time $t$.}
\nomenclature[V]{$q_{s_l,t}$}{Reactive power at sending end $s$ of branch $l$, time $t$.}
\nomenclature[V]{$p_{o_l,t}$}{Active power losses of branch $l$, time $t$.}
\nomenclature[V]{$q_{o_l,t}$}{Reactive power losses of branch $l$, time $t$.}
\nomenclature[V]{$K_{o_l,t}$}{Equivalent ampacity constraint for power losses of branch $l$.}

\nomenclature[P]{$\gamma/\tau$}{Annualization factor, adjusted to the horizon length.}
\nomenclature[P]{$IC_s^{\text{p}}$}{Investment cost per unit rated power of BESS $s$.}
\nomenclature[P]{$IC_s^{\text{e}}$}{Investment cost per unit installed energy of BESS $s$.}
\nomenclature[P]{$OC_{l}$}{Penalty cost associated to branch $l$ system losses.}
\nomenclature[P]{$w_{\text{loss}}$}{Weight for the cost of system losses.}
\nomenclature[P]{$W_s^{\substack{\min\\\max}}$}{Minimum, maximum rated power limits of BESS $s$.}
\nomenclature[P]{$C_s^{\substack{\min\\\max}}$}{Minimum, maximum installed energy limits BESS $s$.}
\nomenclature[P]{{\footnotesize $SoE_s^{\substack{\min\\\max}}$}}{Minimum, maximum state of energy of BESS $s$.}
\nomenclature[P]{$C_{\text{rate}}$}{Battery power C-rate.}
\nomenclature[P]{$w_{\text{slack}}$}{Weight for the cost of slack variables.}
\nomenclature[P]{$t^{i},\>t^f$}{Initial $i$ and final $f$ time steps of each subproblem.}

\nomenclature[S]{$\Omega_1$}{Master problem variables.}
\nomenclature[S]{$\Omega_2$}{Subproblems variables.}
\nomenclature[S]{$\Pi$}{Constraints involving variables in $\Omega_1$.}
\nomenclature[S]{$\Lambda$}{Constraints involving variables in $\Omega_2$.}
\nomenclature[S]{$\Xi$}{Coupling constraints involving variables in $\Omega_1$ and $\Omega_2$.}
\nomenclature[V]{$\Gamma_d^{(\beta)}$}{LHS of optimality cuts.}
\nomenclature[V]{$\Upsilon_d^{(\beta)}$}{LHS of feasibility cuts.}
\nomenclature[S]{$\Phi$}{Feasible subproblems across Benders iterations.}
\nomenclature[S]{$\mathscr{B}$}{Completed Benders iterations.}
\nomenclature[S]{$\mathscr{D}$}{Set of subproblems $d$.}
\nomenclature[V]{$\alpha_{d}$}{Proxy for the operational cost of subproblem $d$.}
\nomenclature[V]{$s_{W_s,t}$}{Slack variable for rated power of BESS $s$.}
\nomenclature[V]{$s_{C_s,t}$}{Slack variable for installed energy of BESS $s$.}

\usepackage[nolist]{acronym}
\begin{acronym}
    \acro{SOCP}{second-order cone programming}
    \acro{SOC}{second-order cone}
    \acro{MISOCP}{mixed-integer second-order cone programming}
    \acro{MIP}{mixed-integer programming}
    \acro{TSO}{transmission system operator}
    \acro{BESS}{battery energy storage system}
    \acro{OPF}{optimal power flow}
    \acro{DC}{direct current}
    \acro{GBD}{Generalized Benders Decomposition}
    \acro{MINLP}{mixed-integer non-linear programming}
    \acro{AC}{alternating current}
    \acro{OPF}{optimal power flow}
    \acro{PF}{power flow}
    \acro{SDP}{semi-definite programming}
    \acro{QCP}{quadratic conic programming}
    \acro{ADMM}{alternating direction method of multipliers}
    \acro{PST}{phase shifting transformer}
    \acro{OLTC}{on-load tap changer}
    \acro{DER}{distributed energy resource}
    \acro{LP}{linear programming}
    \acro{MILP}{mixed-integer linear programming}
    \acro{ADN}{active distribution network}
    \acro{PEGASE}{Pan European Grid Advanced Simulation and state Estimation}
    \acro{DW}{Dantzig-Wolfe}
\end{acronym}

\begin{document}

\title{Enhancing Power Systems Transmission Adequacy via Optimal BESS Siting and Sizing using Benders Decomposition with Feasibility Cuts}
% Operationally driven optimal \acp{BESS} siting and sizing in transmission networks using Benders decomposition with feasibility cuts
\author{Ginevra Larroux, Matthieu Jacobs, Keyu Jia, Fabrizio Sossan~\IEEEmembership{Member,~IEEE}, Mario Paolone,~\IEEEmembership{Fellow,~IEEE} \thanks{The authors are with the École Polytechnique Fédérale de Lausanne (EPFL) and the HES-SO Valais, Switzerland, email: ginevra.larroux@epfl.ch. This work was sponsored by a consortium of industry partners (including Swissgrid) and by Innosuisse, within the ``STORE'' flagship project.}}

\maketitle

\begin{abstract}
This work presents a general framework for the operationally driven optimal siting and sizing of battery energy storage systems in power transmission networks, aimed at enhancing their resource adequacy. The approach considers multi-period planning horizons, enforces network constraints at high temporal resolution, and targets large-scale meshed systems. The resulting computationally complex mixed-integer non-linear programming problem is reformulated as a mixed-integer second-order cone programming problem and solved via Generalized Benders Decomposition, with feasibility cuts enabling congestion management and voltage regulation under binding network limits. A tailored heuristic recovers an alternating-current power-flow-feasible operating point from the relaxed solution. The proposed formulation is parallelizable, yielding excellent computational performance, while featuring rigorous guarantees of convergence. 
\end{abstract}


\vspace{-3mm}
\printnomenclature
\vspace{-3mm}
\section{Introduction}\label{sec:introduction}
Modern power systems with high shares of stochastic generation, increasingly strain system adequacy. The strategic deployment of market-neutral \acp{BESS} by \acp{TSO} can improve the cost-effectiveness of re-dispatch, and reduce the need for grid reinforcement, thereby supporting the economic, safe and reliable system operation. \acp{TSO} therefore increasingly equip themselves with high-granularity market simulations that produce future net-injection trajectories and associated scenario sets. High temporal resolution is required to capture rapid intra-day fluctuations driven by intermittent and weakly dispatchable \acp{DER}, while scenario diversity captures their uncertainty. Seasonal variability in generation and demand is typically accounted for by considering annual or multi-year horizons.

In this context, this work proposes a general framework for the optimal siting and sizing of \acp{BESS} in country-scale meshed transmission networks, trading off \acp{BESS} investment costs against network operational costs (e.g., for losses). Consistent with a \ac{TSO}-driven operational support perspective, network constraints (i.e., the nonlinear \ac{PF} equations) are enforced at hourly resolution over a one-year horizon, and boundary conditions explicitly include ampacity- and voltage-limit violations. \acp{BESS} market neutrality in day-ahead and balancing markets is imposed via hard constraints. The resulting formulation includes integer siting decisions, yielding a \ac{MINLP} that is acknowledged to be computationally intractable for real-world system sizes.

To achieve computational tractability, this work explores potential for parallelization of the problem solution, examines optimization problem decomposition techniques and grid constraints convexification strategies, with emphasis on achieving mathematically rigorous convergence guarantees and ensuring that the adopted relaxations admit tight solutions. 

A primary objective is to ensure that the method can converge to an \ac{AC}-\ac{PF}-feasible operating point. Guarantees of global optimality are desirable but secondary. Finally, for the framework to address congestion management and voltage regulation, the decomposition scheme must explicitly handle operational infeasibilities.

\vspace{-3mm}
\section{Literature review}\label{sec:literature_review}
In line with the above discussed objectives, the literature review focuses on two methodological challenges: (i) accurate modeling of the nonlinear \ac{AC}-\ac{OPF} equations for transmission-level applications, and  (ii) ensuring computational tractability and scalability for large-scale optimization problems. 
\vspace{-3mm}
\subsection{Grid constraints in the optimal power flow problem}
The literature adopts a broad spectrum of grid \ac{OPF} models, ranging from non-approximated and non-convex \ac{AC}-\ac{OPF} formulations, e.g.~\cite{zhao_optimal_2025, PSO} (typically solved via heuristics or gradient-based schemes) to simplified representations such as \ac{DC}-\ac{OPF} approximations, e.g.~\cite{ExpansionPlanning, Near-Optimal}, and linearized \ac{AC}-\ac{OPF} models, e.g.~\cite{massucco_siting_2021, JointDistributionNetwork}, or convex relaxations, in particular \ac{SOC}- and \ac{SDP}-\ac{OPF} formulations, e.g.,~\cite{yi_optimal_2023, SOCP, EnergyStorageSiting}. For applications where physical feasibility and operational reliability are required, numerically robust methods (i.e., methods that reliably converge to a stable, accurate solution under finite-precision arithmetic and small perturbations in inputs, even for ill-conditioned problems~\cite{Higham2002ASNA}) and practical runtimes are essential. Under these criteria, \ac{SOC} relaxations are well suited: they are tight for distribution (radial) systems, often sufficiently accurate for transmission (meshed) networks, and can be solved efficiently using commercial \ac{QCP} solvers~\cite{gurobi, mosek} based on interior-point methods~\cite{AdvancedOptimization}. 

A number of studies apply \ac{SOCP} to storage systems allocation, especially in distribution systems. Works such as~\cite{yi_optimal_2023} aim to achieve dispatchability in \acp{ADN}, incorporate line reinforcement as an alternative to \acp{BESS} deployment, and use Benders decomposition to improve scalability, while \cite{SOCP} targets energy balance and grid support in \acp{ADN}. However, extending these approaches to transmission grids is nontrivial because meshed topologies impose cycle constraints on voltage angles that can compromise relaxation tightness~\cite{Low}. In addition, they omit transmission-relevant flexibility devices such as \acp{PST} and \acp{OLTC} for voltage regulation.

\subsubsection*{On the tightness of the second-order-cone relaxation of the optimal power flow for meshed networks}
Standard formulations of the \ac{SOC} relaxation of the \ac{AC}-\ac{OPF} \cite{Low} do not include bus voltage and branch current angles as problem variables, and they are only reconstructed \emph{a posteriori}, if possible. For meshed networks, however, explicit modeling of bus voltage angles is essential, as they must satisfy the cyclic conditions imposed by closed loops in the network. In \cite{yuan_properties_2020}, bus voltage angles are modeled explicitly, and several properties of the relaxation are established, with emphasis on conditions that promote or guarantee tightness of the resulting operating point. However, the authors argue that these results hold only for \emph{radial} networks and do not extend to \emph{meshed} transmission grids. See Section~\ref{sec:proposed_method} for more details.

\subsubsection*{On the tightness of the second-order-cone relaxation of the optimal power flow for meshed networks}
Standard formulations of the \ac{SOC} relaxation of the \ac{AC}-\ac{OPF}~\cite{Low} do not include bus voltage and branch current angles as variables; they are only reconstructed \emph{a posteriori}, when possible. In meshed networks, however, voltage angles must satisfy cycle constraints induced by network loops. In~\cite{yuan_properties_2020}, voltage angles are modeled explicitly and several properties are established, with emphasis on conditions that promote or guarantee tightness of the resulting operating point. The authors of this work argue that these results apply only to \emph{radial} networks and do not extend to \emph{meshed} transmission grids. Further discussion is provided in Section~\ref{sec:proposed_method}. This limitation does not affect the proposed framework, but it is relevant for \ac{TSO}-oriented applications where \ac{AC}-feasible operating points are required.

Only a limited number of works address transmission networks. For instance,~\cite{IntegratedGeneration} considers generation and storage systems co-planning under a linearized \ac{AC}-\ac{OPF}, while~\cite{Near-Optimal} adopts a \ac{DC}-\ac{OPF} formulation. In both cases transmission constraints are approximated. Other contributions consider operational objectives but do not explicitly account for problem infeasibilities. For example,~\cite{OptimumAllocationofBESS} focuses primarily on power-quality aspects, and the bi-level \ac{SOC}-\ac{OPF} framework in~\cite{EnergyStorageSiting} optimizes \acp{BESS} allocation at the distribution level and evaluates transmission-level impacts via a duality-based method (without enforcing transmission-network constraints). Consequently, they cannot directly resolve \ac{OPF} infeasibilities at the transmission level, e.g. ampacity and voltage limit violations.
\vspace{-3mm}
\subsection{Optimization problem decomposition}
\ac{GBD} is the dominant approach for decomposing optimal assets siting and sizing problems~\cite{yi_dispatch-aware_2023, IntegratedGeneration, chen_optimal_2019}. Alternative decomposition methods for large-scale problems or hierarchical optimization include \ac{DW} decomposition, \ac{ADMM}, and relaxation/cutting-hyperplane methods~\cite{conejo_decomposition_2006}, but they are less explored for operationally driven infrastructure-planning applications. Since \ac{DW} targets complicating constraints rather than complicating variables, it is particularly suitable for problems that decompose across multiple peer substructures (as opposed to a main–subproblem hierarchy), such as multi-area optimal reactive power planning~\cite{DWD} or optimal \ac{BESS} operation~\cite{DWD2}. Similarly, the \ac{ADMM} is widely used for decentralized \ac{OPF} studies in large multi-energy or multi-area networks~\cite{ADMM1, ADMM2}. 
\subsubsection*{On the convergence of Generalized Benders Decomposition for mixed-integer second-order-cone programming problems}
Results from~\cite{benders_partitioning_1962}, \cite{geoffrion_generalized_1972}, and~\cite{SAHINIDIS1991481} establish the standing assumptions for the convergence of the \ac{GBD} algorithm. In particular, with respect to the problem formulation, convergence is guaranteed when either the master problem feasible set is finite (i.a), or a subset of the feasible region of the linking variables in all subproblems (i.b); or the set of cuts generated from the subproblem duals is finite (ii).
For the application under study, condition (i.b) cannot be satisfied because the framework must explicitly handle infeasibility, i.e. violations of branch ampacity and nodal voltage limits. Conditions (i.a) and (ii) require that, respectively, the main problem variables take values from a finite discrete set, or the subproblems be linear, yielding a polyhedral dual feasible set.
Furthermore, convergence of the algorithm requires that optimality and feasibility cuts be derived in accordance with Farkas’ lemma~\cite{geoffrion_generalized_1972}. The work in~\cite{VectorConicConstraint} illustrates their formulation for problems involving convex conic constraints in Banach spaces. 
\subsubsection*{Paper contributions}
This paper develops a computationally tractable formulation for high-temporal-resolution, long-horizon operationally driven planning problems. In view of the above, the proposed paper distinguishes itself from existing literature through the following contributions:
\begin{enumerate}
    \item a \ac{GBD} solution scheme with rigorous convergence guarantees for both optimality and feasibility cuts in the presence of mixed-integer main variables (hence a non-finite main feasible region), enabling siting and sizing under infeasible boundary conditions;
    \item a tailored heuristic to recover a tight solution, i.e., an \ac{AC}-feasible operating point, from the \ac{SOC} relaxation of the \ac{AC}-\ac{OPF} in \textit{meshed} transmission networks;
    \item high computational performance (e.g., relative to~\cite{yi_optimal_2023}), achieved through the parallelization of the subproblem solution.
\end{enumerate}

\vspace{-3mm}
\section{Proposed Method}\label{sec:proposed_method}
In line with established approaches in the literature, the grid constraints in the optimal \acp{BESS} allocation problem (originally a \ac{MINLP}) are relaxed, yielding a \ac{MISOCP} that can be efficiently decomposed using Benders decomposition.
\vspace{-3mm}
\subsection{Optimization problem decomposition with Generalized Benders Decomposition} 
For completeness and to introduce the notation used throughout, we briefly outline the application of Benders decomposition to the considered problem.
The Benders technique decomposes the \ac{MISOCP} into two stages: a main problem and a family of subproblems. Decision variables include the \acp{BESS} site (location) and size (installed energy and rated power), time-invariant variables, together with the time-variant synchronous generators' reactive power setpoints. Other variables include grid state and control variables, notably the active and reactive power and state of energy of \acp{BESS}. 

The \acp{BESS} size variables enter both the investment decisions and the operational grid constraints; therefore serve as the linking variables in the decomposed formulation.

The main problem, formulated as a \ac{MILP}, selects the site and size of \acp{BESS} by minimizing their capital expenditures and a proxy of operational costs aggregated from the subproblems. Each subproblem models one day of grid operation\footnote{A daily decomposition mirrors day-ahead operation and enforces \ac{BESS} state-of-energy continuity over each 24-hour horizon.}, formulated as a \ac{SOC}-\ac{OPF} and minimizes system reactive power losses~\footnote{The \ac{SOC} relaxation of the \ac{OPF} in~\cite{yuan_properties_2020} replaces the branch reactive power balance equality with an inequality~\eqref{eq:q_ol_rel}. See Section~\ref{subsec:tightness} for further details.} conditional on the main decisions. At each iteration, the main proposes updated linking-variable values; the subproblems assess feasibility and optimality and return the corresponding Benders cuts. This iterative exchange continues until the lower and upper bounds from the main and subproblem stages coincide within a prescribed tolerance, at which point a solution consistent with the centralized problem has been obtained. The resulting \ac{GBD} scheme is outlined in Algorithm~\ref{alg:benders_decomp}.
\subsubsection{Mathematical formulation}
The original non-decomposed problem, is formulated as follows
\begin{equation}
\begin{aligned}
    \min_{\Omega_1, \Omega_2} \quad & \frac{\gamma}{\tau} \cdot \text{CAPEX} + \sum_{d \in \mathscr{D}} \text{OPEX}_{d} \\
    \text{s.t.} \quad 
    & \Pi (\Omega_1) \leq 0 \\
    & \Lambda(\Omega_2) \leq 0 \\
    & \Xi(\Omega_1, \Omega_2) \leq 0
\end{aligned}
\end{equation}
\vspace{-3mm}
\begin{subequations}
\begin{align}
    \operatorname{CAPEX} & = \sum_{s \in \mathscr{S}}\left(IC_s^{\text{p}}\cdot W_{s}+IC_s^{\text{e}}\cdot C_{s} \right)  \\     
    \operatorname{OPEX}_{d} & =  \sum_{t \in \mathscr{T}}\left(w_{\text{loss}} \cdot \sum_{l \in \mathscr{L}}OC_{l}\cdot q_{o_l,t}^2\right)
    \end{align}
\end{subequations}
The sets $\Omega_1 = \{U_{s}, W_{s}, C_{s}\}$ and $\Omega_2 = \{V_{n,t}, \theta_{n,t}, p_{n,t}, q_{n,t}, \theta_{l,t}, p_{sl,t}, q_{sl,t}, p_{ol,t}, q_{ol,t}, K_{ol,t}, P_{n_g,t}, \allowbreak Q_{n_g,t}, P_{n_s,t}, Q_{n_s,t}, E_{s,t}\}$ include the main problem and the subproblems variables, respectively. The set $\Pi (\Omega_1)$ includes the site and size constraints for \acp{BESS}\footnote{A lossless battery model is assumed. Losses can be incorporated while preserving convexity of the problem~\cite{batterymodel}.}, 
\begin{subequations}\label{eq:battery_installation_constraints}
    \begin{align}
        W_{s} \in & \left[ W_{s}^{\text{min}}, W_{s}^{\text{max}}\right]\cdot U_s \\
        C_{s} \in & \left[ C_{s}^{\text{min}}, C_{s}^{\text{max}}\right]\cdot U_s \\
        W_{s}\leq & \ C_{s} \cdot C_{\text{rate}}
    \end{align}
\end{subequations} 
The set of constraints $\Lambda(\Omega_2)$ includes the \ac{SOC}-\ac{OPF} grid constraints in \cite{yuan_properties_2020} \eqref{eq:SOC_grid_constraints}~\footnote{For brevity, \ac{PST} and \ac{OLTC} models are omitted; importantly, phase-shift and tap-ratio variables can be included linearly in the \ac{PF} equations.} and \acp{BESS} operational constraints \eqref{eq:bess_oper}. \eqref{eq:recover_feasibility1} constitutes the tightness-promoting constraint, under the hypotheses \eqref{eq:recover_feasibility2}. \eqref{eq:market_neutrality1} and \eqref{eq:market_neutrality2} enforce day-ahead energy and balancing power market neutrality, respectively. For brevity, nodal voltage magnitudes and angles, and generator reactive-power limits, are omitted; all constraints are understood to hold for every index in the corresponding sets. The parameters $A_{nl}^+$, $A_{nl}^-$, $G_n$, $B_n$, $R_l$, and $X_l$ are defined in~\cite{yuan_properties_2020}. The matrices $A_{s_l n}$ and $A_{r_l n}$ are branch-to-node incidence matrices, with entries equal to $1$ at the sending and receiving end of branch $l$, respectively.
\begin{subequations}\label{eq:SOC_grid_constraints}
\begin{flalign}
    & p_{n,t} =\sum_{l\in \mathscr{L}}\left(A_{n l}^{+} p_{s_l,t} - A_{n l}^{-} p_{o_l,t}\right) + G_n V_{n,t}\\ 
    & q_{n,t} = \sum_{l\in \mathscr{L}}\left(A_{n l}^{+} q_{s_l,t} - A_{n l}^{-} q_{o_l,t}\right) - B_n V_{n,t} \\
    & p_{n,t} = P_{n_g,t} - P_{n_s,t} - P_{n_l,t} \\
    & q_{n,t} = Q_{n_g,t} - Q_{n_s,t} - Q_{n_l,t} \\
    & \begin{aligned}
        (A_{s_ln}  - A_{r_ln}) \ V_{n,t} =\;& 2 R_l p_{s_l,t} + 2 X_l q_{s_l,t} \\
                             & - R_l p_{o_l,t} - X_l q_{o_l,t}
    \end{aligned} \\
    & \begin{aligned}\label{eq:2d}
        \theta_{l,t} = & X_l p_{s_l,t} - R_l q_{s_l,t}
    \end{aligned} \\
    & \begin{aligned}\label{eq:theta_l}
        \theta_{l,t} = & (A_{s_ln}  - A_{r_ln}) \ \theta_{n,t} 
    \end{aligned} \\
    & q_{o_l,t} \geq \frac{p_{s_l,t}^2 + q_{s_l,t}^2}{V_{s_l,t}} X_l \label{eq:q_ol_rel} \\
    & K_{o_l,t} \geq q_{o_l,t} \\
    & p_{o_l,t} X_l = q_{o_l,t} R_l \\
    & (A_{s_ln}V_{n,t})( A_{r_ln}V_{n,t}) \sin ^2\left(\theta_l^{\max }\right) \geq \theta_{l,t}^2  \label{eq:recover_feasibility1} \\
    & (\theta_l^{\text{min}}, \theta_l^{\text{max}}) \subseteq (-\frac{\pi}{2},\frac{\pi}{2}), \;\;\theta_l^{\text{min}} = -\theta_l^{\text{max}} \label{eq:recover_feasibility2}
\end{flalign}
\end{subequations}
\vspace{-5mm}
\begin{subequations}\label{eq:bess_oper}
\begin{align}
    & E_{s,t+1} = E_{s,t} + P_{n_s,t} \Delta t\\
    & E_{s,t^i} = E_{s,t^f} \label{eq:market_neutrality1}\\
    & \sum_{s\in \mathscr{S}}P_{s,t} =  0 \label{eq:market_neutrality2}
\end{align}
\end{subequations}
The constraints involving linking variables $\Xi(\Omega_1, \Omega_2)$ include the storage converter capability curve~\footnote{Assuming \acp{BESS} converter P/Q limits are voltage independent.} \eqref{eq:power_converter}, the state-of-energy limits \eqref{eq:soc_dyn1} and continuity boundary conditions \eqref{eq:soc_dyn2}.
\begin{subequations}
\begin{align}
    & W_s^2  \geq P_{n_s,t}^2 + Q_{n_s,t}^2 \label{eq:power_converter} \\
    & E_{s,t} \in \left[SoE^{\min}, SoE^{\max} \right] \cdot C_s \label{eq:soc_dyn1}\\
    & E_{s,t^i} = 0.5 \cdot (SoE^{\max} - SoE^{\min}) \cdot C_s \label{eq:soc_dyn2}
\end{align}
\end{subequations}
\begin{algorithm}[h]
\caption{Generalized Benders Decomposition}
\label{alg:benders_decomp}
\begin{algorithmic}[1]
    \State Initialize $\underline{\alpha}_{d} = 0$, set iteration counter $k =1$.
    \State \textbf{Step 1: main problem} 
    \State Solve $\text{(MP)}$ and compute $LB^{(k)}$.
    \State Pass the optimal linking-variable $\hat{W}_s$,$\hat{C}_s$ values to all subproblems.
    \State \textbf{Step 2: Subproblem stage}
    \ForAll{$d \in \mathscr{D}$}
        \If{subproblem $d$ is feasible}
            \State Solve $\text{(SBP)}_d$ and compute $\Gamma_d^{(k)}$.
            \State Update $\Phi \leftarrow \Phi \cup \{(d,k)\}$.
        \Else
            \State Solve $\text{(FC-SBP)}_d$ and compute $\Upsilon_d^{(k)}$.
        \EndIf
    \EndFor
    \State Compute $UB^{(k)}$.
    \State \textbf{Step 3: Convergence check}
    \If{all $d \in \mathscr{D}$ feasible \textbf{and} $\lvert UB^{(k)} - LB^{(k)} \rvert \>/\> LB^{(k)}< \varepsilon$}
        \State Solve the non-approximated \ac{AC}-\ac{PF}, with as input the optimal operating point.
        \State \textbf{Output:} optimal linking-variables, state variables and total cost from the \ac{PF}.
    \Else
        \State $k \leftarrow k + 1$.
        \State Go to \textbf{Step 1}.
    \EndIf
\end{algorithmic}
\end{algorithm}
When decomposed, the main problem takes the form
\begin{subequations}
\begin{align}
    \text{(MP)}: & \min_{\Omega_1, \alpha_d} \frac{1}{\tau} \cdot \text{CAPEX} + \sum_{d \in \mathscr{D}} \alpha_d \\
    & \text{s.t.} \notag\\
    & \Pi(\Omega_1) \leq 0 \\
    & \alpha_d \geq \underline{\alpha}, \forall d \in \mathscr{D} \label{eq:init} \\
    & \alpha_d \geq \Gamma^{(\beta)}_d, \forall (d,\beta) \in \Phi \label{eq:opt_cuts} \\
    & \Upsilon_d^{(\beta)} \leq 0, \forall (d, \beta) \in (\mathscr{D} \times \mathscr{B}) \setminus \Phi \label{eq:feas_cuts}
\end{align}
\end{subequations}
with  $\alpha_{d}$ proxy variable for the operational cost of subproblem $d \in \mathscr{D}$, $\mathscr{B} = \{1,\dots,k{-}1\}$ set of completed Benders iterations, with $k$ denoting the current iteration and $\Phi := \left\{(d,\beta) \in \mathscr{D} \times \mathscr{B} \,\middle|\, \text{subproblem } d \text{ is feasible at iteration } \beta \right\}$ set of feasible subproblems across Benders iterations. As the algorithm progresses, optimality cuts \eqref{eq:opt_cuts} and feasibility cuts \eqref{eq:feas_cuts} are iteratively added to the set of constraints. The right-hand-sides of these linear inequalities take the form \eqref{eq:benders_cuts} and \eqref{eq:benders_feas_cuts}, and are derived from the subproblem \eqref{eq:sbp} and the feasibility check subproblem \eqref{eq:feas_check_sbp}, respectively.
\begin{subequations}\label{eq:sbp}
\begin{align}
    \text{(SBP)}_d : & \min_{\Omega_1, \Omega_2} \quad  \text{OPEX}_d \label{eq:subprob_obj} \\
    & \text{s.t.} \notag\\
    & \Lambda(\Omega_2) \leq 0 \label{eq:subprob_lambda} \\
    &  \Xi(\Omega_1, \Omega_2) \leq 0 \label{eq:subprob_xi} \\
    &  W_s = \hat{W}_s \quad :  \lambda_{s,d} \label{eq:subprob_lambda_dual} \\
    &  C_s = \hat{C}_s \quad :  \mu_{s,d} \label{eq:subprob_mu_dual}
\end{align}
\end{subequations}
\vspace{-2mm}
\begin{equation}\label{eq:benders_cuts}
\begin{aligned}
    \Gamma_d^{(\beta)} & = \overline{\text{OPEX}}_d^{(\beta)} + \sum_{s \in \mathscr{S}} \bar{\lambda}_{s,d}^{(\beta)}\left(W_s - \hat{W}_s^{(\beta)}\right) + \\
    & \sum_{s \in \mathscr{S}} \bar{\mu}_{s,d}^{(\beta)}\left(C_s - \hat{C}_s^{ (\beta)}\right)
\end{aligned}
\end{equation}
\vspace{-2mm}
\begin{subequations}\label{eq:feas_check_sbp}
\begin{align}
    \text{(FC-SBP)}_d:& \min_{\Omega_1, \Omega_2, s_{W_s}, s_{C_s}}  w_{\text{slack}} \sum_{s \in \mathscr{S}} (s_{W_s} + s_{C_s})\\
    & \text{s.t.} \notag\\
    & \Lambda(\Omega_2) \leq 0 \\
    & \Xi(\Omega_1, \Omega_2) \leq 0  \\
    & W_s = \hat{W}_s + s_{W_s}  \quad :\nu_{s,d} \label{eq:link_dual_W_s}\\
    & C_s = \hat{C}_s + s_{C_s}  \quad :\xi_{s,d} \label{eq:link_dual_C_s} \\
    & s_{W_s}, s_{C_s} \geq 0
\end{align}
\end{subequations}
\vspace{-2mm}
\begin{equation}\label{eq:benders_feas_cuts}
\begin{aligned}
    \Upsilon_d^{(\beta)} & = w_{\text{slack}} \sum_{s \in \mathscr{S}} \left(\tilde{s}_{W_s,d}^{(\beta)} + \tilde{s}_{C_s,d}^{(\beta)} \right)  + \\ 
    &\sum_{s \in \mathscr{S}} \tilde{\nu}_{s,d}^{(\beta)}\left(W_s - \hat{W}_s^{ (\beta)}\right)+\sum_{s \in \mathscr{S}} \tilde{\xi}_{s,d}^{(\beta)}\left(C_s - \hat{C}_s^{(\beta)}\right)
\end{aligned}
\end{equation}
Throughout this formulation, the operator $\hat{(\cdot)}$ denotes the optimal variables output of the main problem, $\bar{(\cdot)}$ of the subproblems, and $\tilde{(\cdot)}$ of the feasibility check subproblems.

The lower bound on the total cost is computed as $LB ^{(k)}= \frac{\gamma}{\tau} \hat{\operatorname{CAPEX}}^{(k)} + \sum_{d\in\mathscr{D}}\hat{\alpha}_d^{ (k)}$, $k$ current Benders iteration, and its upper bound as $UB ^{(k)}= \frac{\gamma}{\tau} \hat{\operatorname{CAPEX}}^{(k)} + \sum_{d\in\mathscr{D}}\overline{\operatorname{OPEX}}_d^{(k)}$. 
\subsubsection{Computational complexity and parallel solution approach}\label{subsubsec:computational_complexity}
The optimization framework is implemented in CVXPY~\cite{agrawal2018rewriting} and solved using the MOSEK~\cite{mosek} conic optimizer. Each subproblem is constructed with parametrized linking variables and ``dummy-solved'' once to trigger compilation and caching.\footnote{The first \texttt{solve} call checks DCP compliance, canonicalizes the model into a cone program, builds the solver data, and caches the compiled representation.} This initialization stage is executed sequentially on a single process. Afterwards, at each iteration, subproblems are distributed across all available CPU cores and solved in parallel. Model construction is decoupled from model evaluation\footnote{At each iteration, only the coefficient entries affected by the linking parameters are updated before the conic solver is called.}, to keep memory usage bounded and substantially reduce total run time.
\begin{table}[H]
    \centering
    \caption{Centralized problem size.}
    \scriptsize
    \setlength{\tabcolsep}{4pt}
    \renewcommand{\arraystretch}{1.12}
    \begin{tabular}{lc}
        \toprule
        & \textbf{Centralized} \\
        \midrule
        Var (bin)     & $S$ \\
        Var (cont) & $(2 + G + 4N + 6L)T + (2T + (T+1) + 2)S$ \\
        Cons            & $(7 + 2G + 8N + 10L)T + (2T + 2(T+1) + 6 + 2D)S$ \\
        \bottomrule
    \end{tabular}
    \label{tab:computational_complexity_cen}
\end{table}
\vspace{-5mm}
\begin{table}[H]
    \centering
    \caption{Decomposed problem size.}
    \scriptsize
    \setlength{\tabcolsep}{4pt}
    \renewcommand{\arraystretch}{1.12}
    \resizebox{\linewidth}{!}{
    \begin{tabular}{lcc}
        \toprule
        & \textbf{MP} & \textbf{SBP} \\
        \midrule
        Var (bin)     & $S$ & $0$ \\
        Var (cont) & $2S + D$ & $(2 + G + 4N + 6L)P + (2P + (P+1) + 2)S$ \\
        Cons          & $5S + D(|\mathscr{B}|+1)$ & $(7 + 2G + 8N + 10L)P + (2P + 2(P+1) + 5)S$ \\
        Par            & $0$ & $2S$ \\
        \bottomrule
    \end{tabular}}
    \label{tab:computational_complexity_dec}
\end{table}
\vspace{-5mm}
\begin{table}[H]
    \centering
    \caption{Centralized vs. decomposed problem sizes, for the IEEE 118-bus system.}
    \scriptsize
    \setlength{\tabcolsep}{4pt}
    \renewcommand{\arraystretch}{1.12}
    \begin{tabular}{lccc}
        \toprule
        & \textbf{Centralized} & \multicolumn{2}{c}{\textbf{Decomposed}} \\
        \cmidrule(lr){3-4}
        &  & \textbf{MP} & \textbf{SBP} \\
        \midrule
        Var (bin)     & $118$      & $118$          & $0$     \\
        Var (cont) & $17126154$ & $260$          & $47274$ \\
        Cons            & $29161524$ & $590 + 24(|\mathscr{B}|+1)$ & $80482$ \\
        Par            & $0$        & $0$            & $236$   \\
        \bottomrule
    \end{tabular}
    \label{tab:computational_complexity_numbers}
\end{table}
The sizes of the centralized and of the decomposed formulation are reported in Table~\ref{tab:computational_complexity_cen} and Table~\ref{tab:computational_complexity_dec}, respectively\footnote{$G:=|\mathscr{G}|$, $N:=|\mathscr{N}|$, $L:=|\mathscr{L}|$, $S:=|\mathscr{S}|$, $T:=|\mathscr{T}|$, $D:=|\mathscr{D}|$, $P:=|\mathscr{T}|/|\mathscr{D}|$.}\footnote{``Var (bin)'' and ``Var (cont)'' denote binary and continuous variables; ``Par'' denotes parameters and ``Cons'' constraints, following CVXPY’s terminology.}. The subproblem size is reported per instance; total counts in Table~\ref{tab:computational_complexity_dec} and \ref{tab:computational_complexity_numbers}, column ``SBP'' scale with the number of subproblems (or, under parallel execution, with the number of concurrently solved instances). Notably, subproblem dimensions remain constant across Benders iterations. In contrast, the main-problem size grows linearly with the iteration count $|\mathscr{B}|$, since one cut per subproblem is generated and retained at each iteration.

Notably, the subproblem dimensions remain constant across Benders iterations. In contrast, the main problem size grows linearly with the iteration count $|\mathscr{B}|$, since one cut per subproblem is generated and retained at each Benders iteration. For the IEEE 118-bus system (namely, the test network used in the case study) over a one-year horizon, the corresponding numerical sizes are reported in Table~\ref{tab:computational_complexity_numbers}. 
\subsubsection{Convergence}
The work in~\cite{VectorConicConstraint} extends the \ac{GBD} framework to \ac{MINLP} with vector conic constraints in Banach spaces, providing the theoretical basis for optimality and feasibility cuts formulation through a generalized version of Farkas’ lemma. The geometric interpretation of the lemma is that a vector either lies in the cone or can be strictly separated from it by a hyperplane. In \ac{GBD}, this result provides a certificate of infeasibility and underpins the formulation of the feasibility-check subproblem, that is a numerical procedure to compute such a separating hyperplane. 

In the present setting, the hypothesis in~\cite{geoffrion_generalized_1972} that the main problem feasible set be finite is not satisfied, since the main problem includes both (site) binary variables and (size) continuous variables. Nevertheless, convergence guarantees remain valid; see Appendix~\ref{sec:appendix} for a formal argument.

Moreover, unlike existing formulations (e.g.,~\cite{shahidehpour_benders_2005, VectorConicConstraint}), slack variables are introduced in the feasibility-check subproblem \emph{only} for the linking variables \eqref{eq:link_dual_W_s}, \eqref{eq:link_dual_C_s}, rather than for every constraint. This remains consistent with Farkas’ lemma provided that the slack variables enter the conic constraints linearly, thereby preserving the theoretical convergence guarantees of \ac{GBD}. A formal justification is provided in Appendix~\ref{sec:appendix}.

Finally, note that the subproblems \eqref{eq:sbp} solve convex relaxations of the original \ac{AC}-\ac{OPF}. In the feasibility-check subproblem, the Farkas’ lemma requires an objective that penalizes only the slack variables (losses cannot be included to promote tightness, as is common in standard \ac{SOC}-\ac{OPF} formulations). Consequently, the feasibility-check subproblem might return an operating point that differs from the regular subproblem and slow convergence: part of the infeasibility may be absorbed through the artificially inflated losses permitted by the relaxation, rather than being reflected solely in the slack variables. Nevertheless, the algorithm’s convergence guarantees remain unaffected.
\vspace{-3mm}
\subsection{Second-order cone relaxation of the optimal power flow}
\subsubsection*{Tightness}
The \ac{SOC}-\ac{OPF} model adopted here follows \cite{yuan_properties_2020}, which explicitly models bus voltage angles and advocates applicability to both radial and meshed topologies. However, the authors of this paper argue that tightness cannot be formally guaranteed for meshed networks. This issue is discussed in further detail in another work by the authors~\cite{letter}, where this limitation is formally established and supported by experimental evidence showing that, even with a numerically zero total relaxation gap, substituting the relaxed solution into the nonlinear \ac{AC}-\ac{PF} equations yields nonzero residuals on IEEE test networks. 

Briefly, constraint~\eqref{eq:2d} in the \ac{SOC}-\ac{OPF} formulation of \cite{yuan_properties_2020} is derived by approximating the \ac{AC}-\ac{OPF}-feasible equation~\eqref{eq:1e},
\begin{equation}\label{eq:1e}
    v_{sl} v_{rl} \sin(\theta_{l,t}) = X_l p_{sl,t} - R_l q_{sl,t},
\end{equation}
and the cyclic constraint~\eqref{eq:cyclic_con} is claimed to hold when~\eqref{eq:theta_l} is included in the set of constraints
\begin{equation}\label{eq:cyclic_con}
    \sum_{l \in \mathscr{C}} \theta_l
    = 0 \bmod 2\pi, \qquad \forall\mathscr{C} \text{ cycle}.
\end{equation}
Lemma~1 in~\cite{letter} shows that this implication does not hold in general. Moreover, the monotonic load-increase procedure required to promote tightness in Theorems~5 and~6 of~\cite{yuan_properties_2020} is not admissible in the in the context of this work, since load injections are fixed parameters. For this reason, a semi-heuristic procedure is adopted to recover an \ac{AC}-\ac{PF}-feasible solution to the planning problem. Specifically, once the algorithm has converged, an \ac{AC}-\ac{PF} is solved using the state variables obtained from the \ac{SOC}-\ac{OPF} (see Stage~17 in Algorithm~\ref{alg:benders_decomp}).
\vspace{-3mm}
\section{Case studies and results}\label{sec:experimental_validation}
The simulations are carried out on a MacBook Pro equipped with an Apple M4 Pro chip featuring $10$ efficiency cores and \SI{48}{GB} of RAM.

The IEEE 118-bus network~\cite{ieee118} is used as test system. Time-series profiles (hourly) of nodal active power net injections (generation minus load) are sourced from IEEE DataPort\footnote{\url{https://ieee-dataport.org/documents/time-series-data-load-generation-and-power-flow-ieee-bus-systems}}. These injections are derived from a \ac{DC}-\ac{PF}, and therefore neglect reactive power and losses. To obtain physically meaningful active and reactive injections for the planning problem, an \ac{AC}-\ac{PF} initialization is performed. Specifically, load power factors are computed for all PQ buses and for buses hosting only synchronous condensers using the snapshot operating-point values in the MATPOWER case file\footnote{\url{https://matpower.org/docs/ref/matpower5.0/case118.html}}. Reactive injections at these buses are then reconstructed from the time-series active power and the corresponding power factors. For synchronous generator buses, the time-series data does not distinguish load from generation; therefore, the reactive component of the local load is approximated using the load profile of the PQ bus whose power factor is closest to the nominal MATPOWER value.

\begin{figure*}[h]
    \centering
    \includegraphics[width=\textwidth]{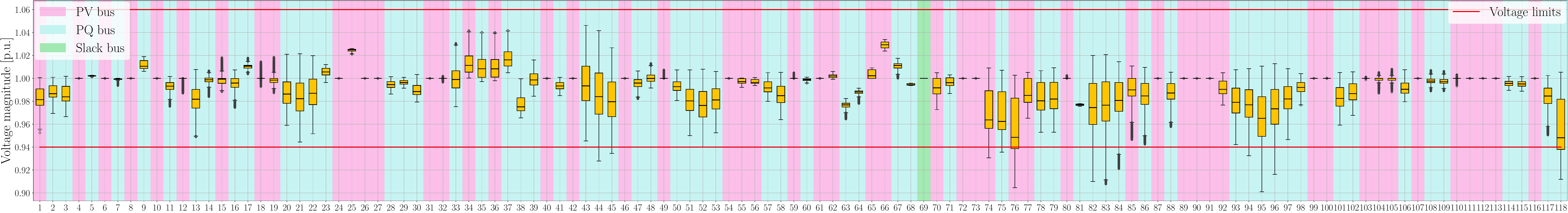}\\[0.6em]
    \includegraphics[width=\textwidth]{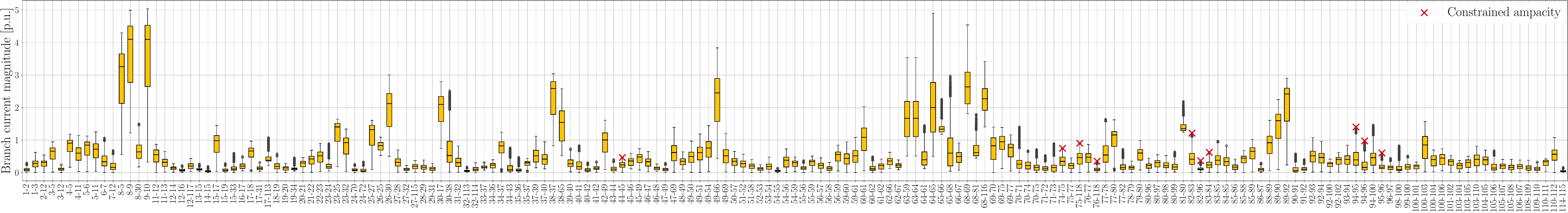}
    \vspace{-5mm}
    \caption{Boundary conditions: distribution of voltage (top) and branch current (bottom) magnitude over the time horizon, per bus and branch respectively.}
    \label{fig:bc}
\end{figure*}

The \ac{AC}-\ac{PF} is then solved using the reconstructed \(P\)--\(Q\) injections, a flat-start (i.e. nodal voltage magnitudes set to $1$, and phases to $0$) initialization for PV-bus voltages, and reactive power limits from the MATPOWER case file. Over the selected time horizon, the resulting operating points exhibit voltage-magnitude violations at twelve buses, and the slack generator exceeds both active and reactive power limits. These injections serve as input to the optimal \ac{BESS} allocation problem, which seeks to resolve voltage and ampacity violations. To explicitly induce congestion, additional infeasibilities are introduced by tightening selected line ampacity limits. Specifically, buses with voltage violations are selected as candidate battery locations and the ampacity limits of lines incident to these buses are reduced so that the network becomes infeasible in the absence of storage. In contrast, slack-bus active power limits are relaxed, since this infeasibility cannot be addressed by storage constrained to have equal state of energy at the beginning and end of each day. These boundary conditions are illustrated in Figure~\ref{fig:bc}.
Before evaluating the proposed framework for the optimal \ac{BESS} allocation problem, three preliminary studies are performed: (i) verification of \ac{GBD} convergence to the centralized solution, (ii) analysis of \ac{GBD} solve-time scalability with respect to the planning horizon and the number of candidate battery locations, and (iii) assessment of the tightness of the relaxed grid constraints.
\vspace{-3mm}
\subsection{Generalized Benders Decomposition convergence}
The planning problem is a non-convex \ac{MISOCP}. Consequently, neither the centralized formulation nor its Benders-decomposed counterpart provides guarantees of global optimality. To verify the correctness of the decomposed formulation, two sets of boundary conditions, tailored to isolate the generation of (1) optimality and (2) feasibility cuts, are considered.

To test the optimality-cut mechanism, the binary variables are relaxed, obtaining a centralized convex \ac{SOCP}. In this setting, \ac{GBD} is guaranteed to converge to the global optimum of the relaxation and thus to the centralized solution. All operational infeasibilities are removed\footnote{Voltage-magnitude limits and slack-bus active power constraints are relaxed.} to ensure that subproblems generate optimality cuts only. To validate feasibility cuts, the finite-convergence assumptions in~\cite{geoffrion_generalized_1972} requiring polyhedral subproblem feasible sets are exploited. The \ac{SOC}-\ac{OPF} grid constraints in the subproblems are therefore replaced with a \ac{DC}-\ac{OPF}, yielding \ac{LP} subproblems, which combined with a \ac{LP} main problem, produce a centralized \ac{LP}. In this second setting, infeasibilities in the boundary conditions (voltage violations and line congestions) are included so that feasibility cuts are generated.

\begin{figure}[h]
    \centering
    \includegraphics[width=0.48\linewidth]{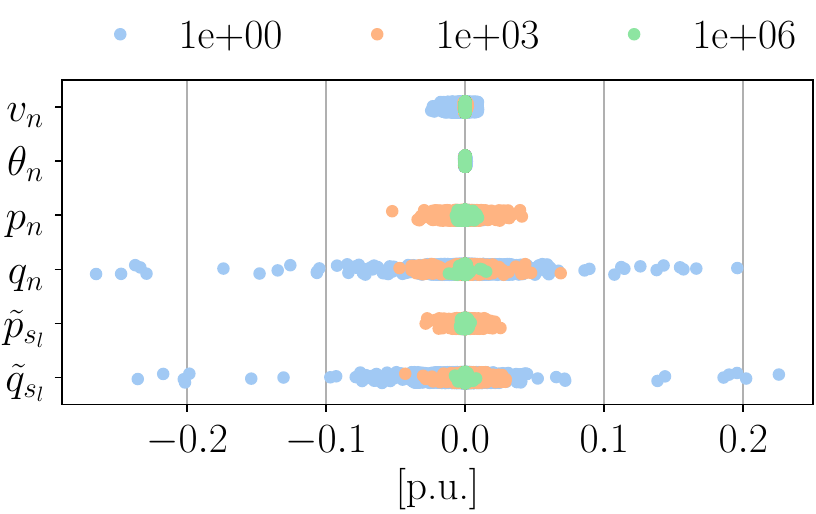}
    \hfill
    \includegraphics[width=0.48\linewidth]{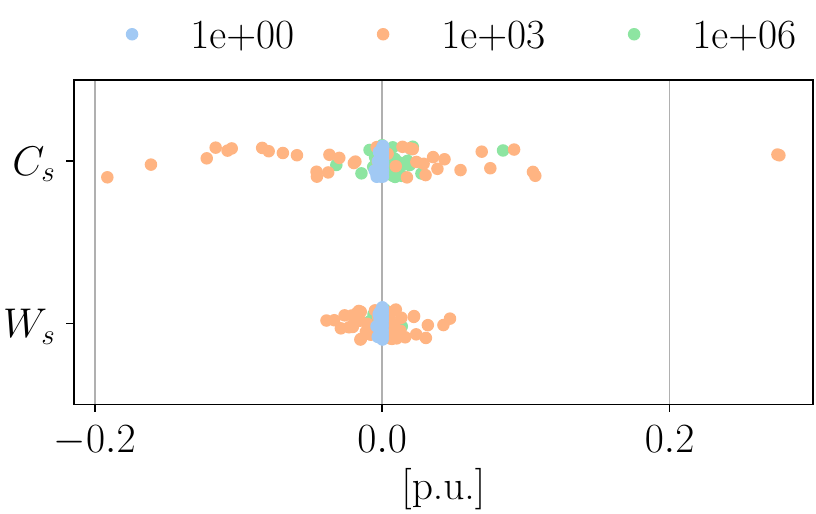}
    \caption{Residuals between the optimal solutions of the centralized and Benders-decomposed integrality-relaxed problem: feasible setting. \ac{SOC}-\ac{OPF} state variables (left) and linking variables (right).}
    \label{fig:centralized_vs_decomposed_feasible}
\end{figure}

\begin{figure}[h]
    \centering
    \includegraphics[width=0.48\linewidth]{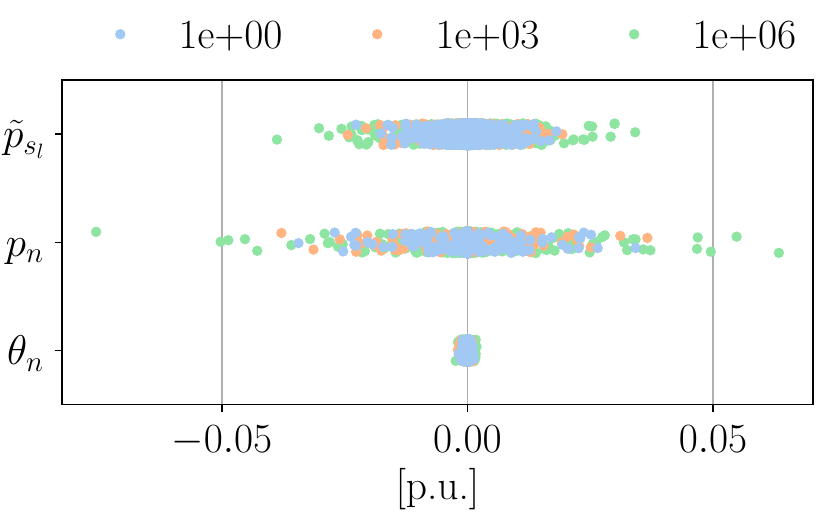}
    \hfill
    \includegraphics[width=0.48\linewidth]{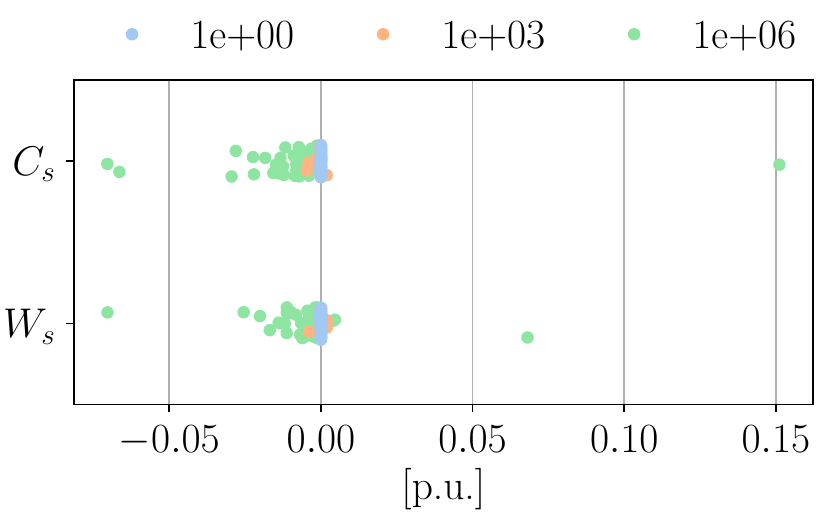}
    \caption{Residuals between the optimal solutions of the centralized and Benders-decomposed integrality-relaxed problem: infeasible setting. \ac{DC}-\ac{OPF} state variables (left) and linking variables (right).}
    \label{fig:centralized_vs_decomposed_infeasible}
\end{figure}

Figure~\ref{fig:centralized_vs_decomposed_feasible} reports the residuals of state variables~\footnote{$\tilde p_{s_l}$, $\tilde q_{s_l}$ are the measurable active and reactive power at the sending end $s$ of branch $l$, see \cite{yuan_properties_2020} for further details.} (left) and linking variables (right) between the centralized and decomposed solutions in the feasible setting, for four representative days\footnote{The four representative days are selected based on the seasonal variation of the daily active power profile of the PQ bus with the largest load. Daily profiles are clustered by season (Dec--Jan--Feb, Mar--Apr--May, Jun--Jul--Aug, Sep--Oct--Nov), and the day with minimum RMSE with respect to the seasonal median is chosen as representative.} and for different weights on power losses in the subproblem objective. The \ac{GBD} termination criterion is set to $(UB-LB) < \delta$ with $\delta = 1\mathrm{e}{-}1$~p.u. (instead of Stage~16 in Algorithm~\ref{alg:benders_decomp}) to control the accuracy with respect to the linking-variable decisions and to avoid an explicit dependence on the loss weight magnitude. The corresponding value of $\varepsilon$ in Stage~16 of Algorithm~\ref{alg:benders_decomp}, computed \emph{a posteriori}, is $1\mathrm{e}{-}2$, $1\mathrm{e}{-}4$, and $1\mathrm{e}{-}6$ for the three tested loss weights, respectively. Two opposite trends are observed: state-variable residuals are largest for small loss weights, whereas linking-variable residuals are smallest. As the loss weight increases, the operational term dominates the objective and the relative contribution of the size variables becomes marginal; consequently, for large loss weights, even if state-variable residuals are negligible, indicating matching operating points, linking-variable residuals approach the termination tolerance. Overall, the near-zero (or tolerance-level) residuals obtained for $C_s$ and $W_s$ for a unitary loss weight support the correct implementation of the optimality cuts within \ac{GBD}. 
Analogously, Figure~\ref{fig:centralized_vs_decomposed_infeasible} reports the residuals of the same quantities for the infeasible boundary conditions. Here, the variable weight is on the slack bus power in the subproblem objective. The termination threshold is set to $\delta = 1\mathrm{e}{-}2$, which leads to $\varepsilon = 1\mathrm{e}{-}8$. Again, near-zero (or tolerance-level) residuals of the linking variables for small subproblem objective weights support the correct implementation of feasibility cuts. 
\vspace{-3mm}
\subsection{Solution time and scalability}
The next simulations evaluate the computational performance of the \ac{GBD} algorithm applied to the non-convex \ac{MISOCP}. Here, ``convergence'' refers to termination of the algorithm in a finite number of iterations; it does not imply global optimality, nor equality to the centralized solution.

\begin{figure}[h]
    \centering
    \includegraphics[width=0.48\linewidth]{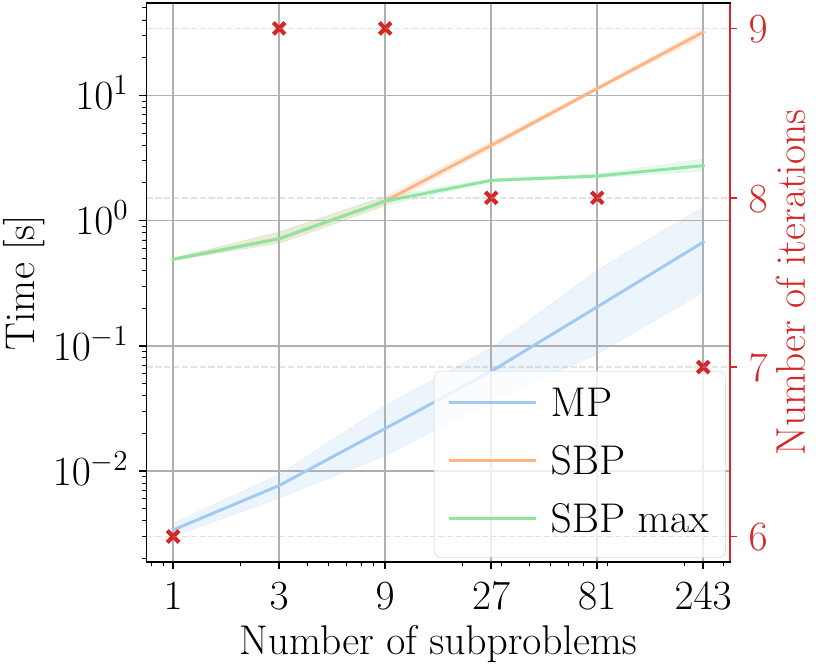}
    \hfill
    \includegraphics[width=0.48\linewidth]{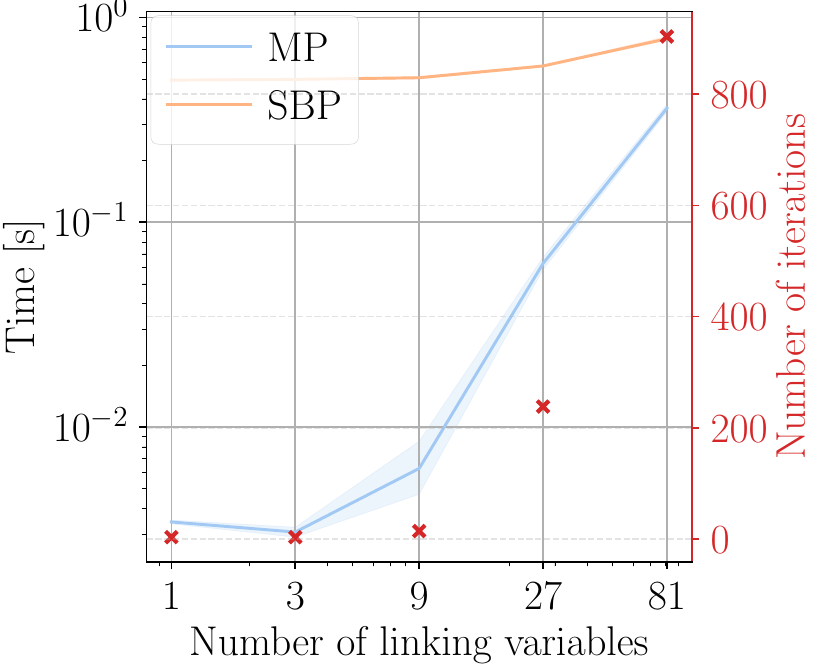}
    \caption{Main-problem solve time per iteration (blue), total subproblem solve time per iteration (orange), maximum subproblem solve time across workers (green) and number of Benders iterations (red) as functions of the number of subproblems (left) and number of linking variables (right). Mean (solid line) and standard deviation (shaded area) across iterations.}
    \label{fig:scalability}
\end{figure}

Figure~\ref{fig:scalability} illustrates the scalability of the proposed framework with respect to the number of subproblems (i.e., the planning horizon length) (left) and the number of linking variables (battery size variables) (right), both varied geometrically. The termination tolerance is set to $\varepsilon=$ $1\mathrm{e}{-}4$ and $1\mathrm{e}{-}2$, respectively.

\begin{figure}[h]
    \centering
    \includegraphics[width=0.48\linewidth]{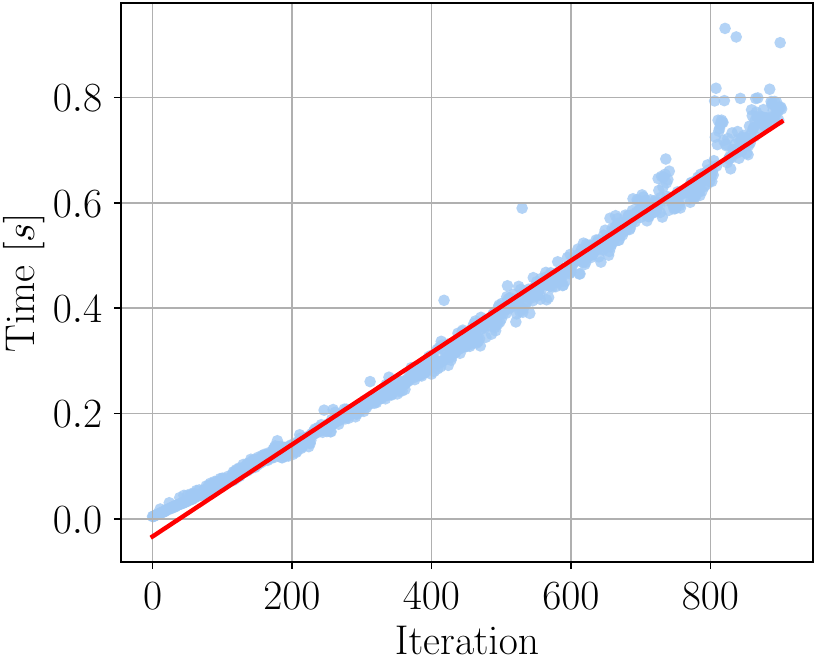}
    \hfill
    \includegraphics[width=0.48\linewidth]{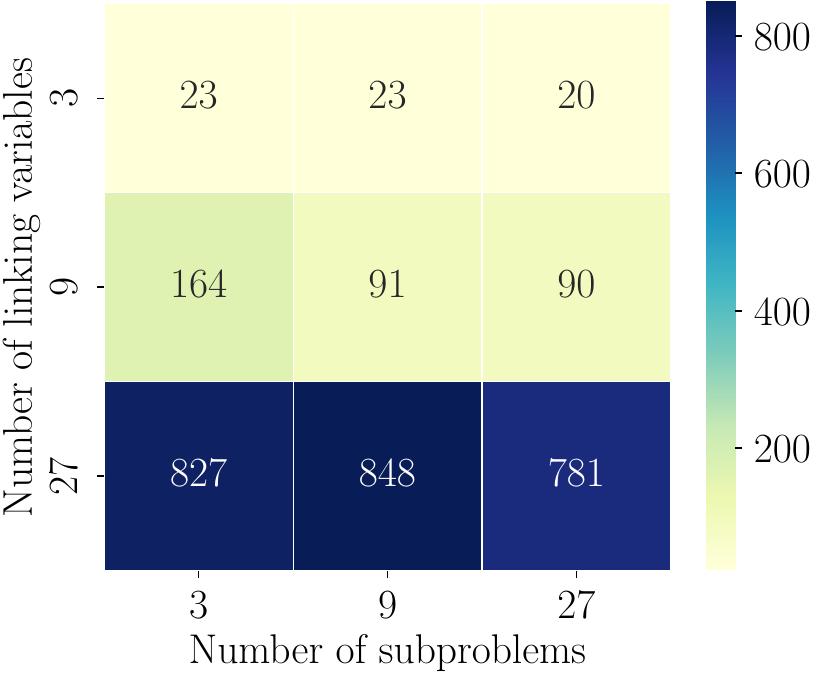}
    \caption{Main problem solve-time per iteration as a function of Benders iteration count (left) and number of Benders iterations as a joint function of number of subproblems and linking variables (right).}
    \label{fig:scalability_specs}
\end{figure}

Concerning the number of subproblems, consistently with the discussion on computational complexity in Section~\ref{subsubsec:computational_complexity}, the main problem solve time exhibits an approximately linear dependence on the planning horizon length~\footnote{A log--log linear fit of the form $\log_{10} y = m\log_{10} x + q$ yields $m \approx 0.9$, i.e., close to the ideal linear scaling ($m=1$), and $10^q\approx 0$.}. Its dispersion across iterations is non-negligible and is further evidenced in Figure~\ref{fig:scalability_specs} (left), which shows an increasing main problem solve time as a function of the Benders iteration index due to the accumulation of cuts. For the subproblems, the cumulative per-iteration solve time grows linearly with the number of subproblems (for number of subproblems $\geq 9$), with a proportionality constant close to the number of available parallel workers (CPU cores), whereas the per-iteration maximum solve time across workers plateaus. Finally, the number of iterations required for termination remains approximately constant or decreases (cf.\ Figure~\ref{fig:scalability_specs} (right)), which is consistent with the fact that increasing the number of subproblems yields more cuts per iteration and therefore restricts the main-feasible region more rapidly.

Concerning the number of linking variables, the main problem solve time increases more than linearly, as the main is a \ac{MILP} whose integer dimension scales with the number of candidate locations. In contrast, the subproblem solve time per iteration remains approximately constant. The number of Benders iterations grows rapidly with the number of linking variables, indicating that the algorithmic effort required for termination increases markedly as the dimension of the binary and linking-variable space grows. This constitutes the main practical limitation of the approach. A closed-form scaling law is not reported, as the observed iteration growth is strongly affected by the spatial distribution of candidate locations.

Overall, Table~\ref{tab:scalability_num_sbps_totalsolvetime} and Table~\ref{tab:scalability_num_sbps_iter} report the total solution time for the main problem and subproblems at termination, confirming the computational efficiency of the approach. The dominant contributors to overall runtime are, in decreasing order, the number of linking variables, the termination tolerance (which drives the iteration count), and the number of subproblems, which is partly mitigated by parallelization. In practice, this suggests that long planning horizons are computationally tractable, whereas the number of candidate \ac{BESS} locations should be kept moderate. 
\begin{table}[H]
    \centering
    \caption{Main (MP) and subproblem (SBP) total solution time as a function of the number of subproblems.}
    \label{tab:scalability_num_sbps_totalsolvetime}
    \scriptsize
    \setlength{\tabcolsep}{5pt}
    \renewcommand{\arraystretch}{1.12}
    \resizebox{\linewidth}{!}{
    \begin{tabular}{lccccccc}
        \toprule
        \textbf{\# days} 
        & \textbf{1} & \textbf{3} & \textbf{9} & \textbf{27} 
        & \textbf{81} & \textbf{243} & \textbf{365} \\
        \midrule
        \textbf{MP}   [\SI{}{\second}]
        & $0.05$ & $0.07$ & $0.18$ & $0.45$ & $1.44$ & $4.05$ & $7.35$ \\
        \textbf{SBP} [\SI{}{\second}]
        & $2.94$ & $17.94$ & $108.80$ & $293.69$ & $867.69$ & $2193.81$ & $5192.92$ \\
        \bottomrule
    \end{tabular}}
\end{table}
\vspace{-5mm}
\begin{table}[H]
    \centering
    \caption{Main (MP) and subproblem (SBP) total solution time as a function of the number of linking variables.}
    \label{tab:scalability_num_sbps_iter}
    \scriptsize
    \setlength{\tabcolsep}{5pt}
    \renewcommand{\arraystretch}{1.12}
    \resizebox{\linewidth}{!}{
    \begin{tabular}{lcccccc}
        \toprule
        \textbf{\# linking variables} 
        & \textbf{1} & \textbf{3} & \textbf{9} & \textbf{27} 
        & \textbf{81} & \textbf{118}  \\
        \midrule
        \textbf{MP}   [\SI{}{\second}]
        & $0.04$ & $0.02$ & $0.09$ & $14.91$  & $325.34$ &  $3432.16$ \\
        \textbf{SBP} [\SI{}{\second}]
        & $1.48$ & $1.50$ & $7.11$ & $137.86$ & $711.21$ & $1972.77$  \\
        \bottomrule
    \end{tabular}}
\end{table}
\vspace{-3mm}
\subsection{Tightness and AC-power flow feasibility}\label{subsec:tightness}
\begin{figure}[h]
    \centering
    \includegraphics[width=0.48\linewidth]{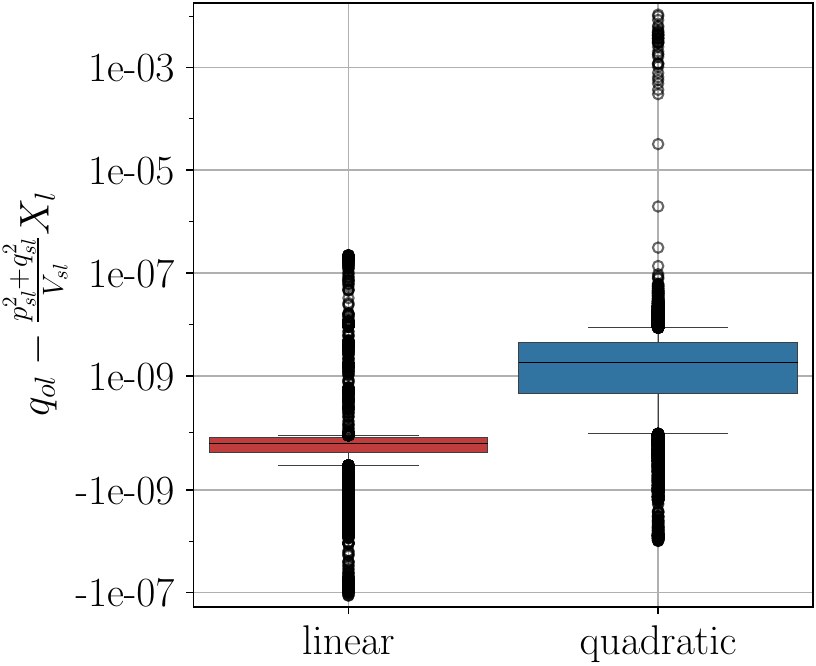}
    \includegraphics[width=0.48\linewidth]{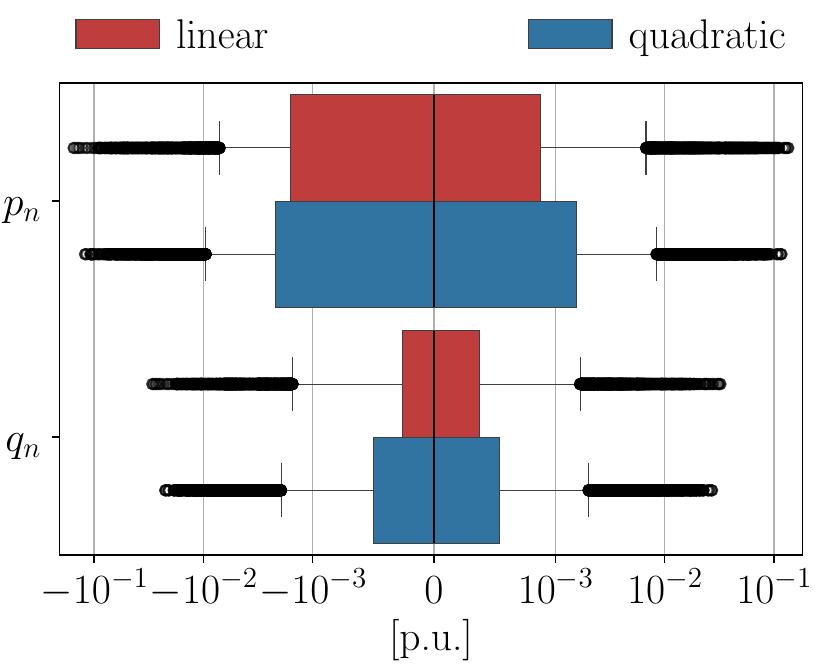}
    \caption{Distribution of the relaxation gap on reactive power losses over the time horizon (left), and nodal power residuals between the optimization solution and the \ac{AC}-\ac{PF} equations evaluated at the optimized voltage phasors (right), for linear (red) and quadratic (blue) loss costs.}
    \label{fig:linear_vs_quadratic_tightness}
\end{figure}

\begin{figure}[h]
    \centering
    \includegraphics[width=0.6\linewidth]{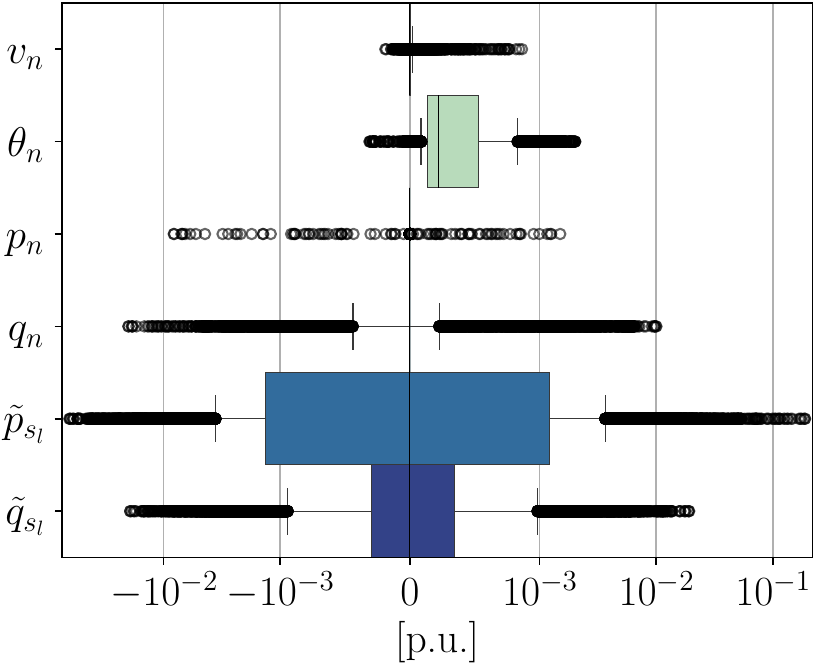}
    \caption{Residuals between the optimization solution and the \ac{AC}-\ac{PF} solved using the optimization solution as initialization.}
    \label{fig:quadratic_tightness}
\end{figure}

Tightness of the \ac{SOC}-\ac{OPF} relaxation is commonly assessed via the relaxation gap. A zero gap is not generally guaranteed when loads are fixed, and the aggregate gap magnitude alone is not a reliable metric of proximity to an \ac{AC}-\ac{PF}-feasible operating point. As shown in Figure~\ref{fig:linear_vs_quadratic_tightness} (left), the relaxation gap obtained with a linear loss penalty is very small (on the order of $1\mathrm{e}{-}6\div1\mathrm{e}{-}7$ p.u.); nonetheless, the nodal active/reactive power residuals in Figure~\ref{fig:linear_vs_quadratic_tightness} (right) (obtained by evaluating the nonlinear \ac{AC}-\ac{PF} equations at the optimized voltage magnitudes and angles) indicate that the resulting operating point is not feasible. Under a linear loss penalty, the gap tends to concentrate on a limited subset of branches, where artificially inflated losses incur the lowest marginal cost. This may violate cycle constraints on voltage phase angles and thereby yield \ac{AC}-infeasible solutions.
To partially mitigate this effect, a more distributed penalty is considered, implemented as a quadratic cost on reactive power losses. While this objective spreads the gap more evenly across branches (as reflected by a wider interquartile range and outliers reaching $1\mathrm{e}{-}2$~[p.u.]), it provides only limited improvements in \ac{AC} consistency: the maximum nodal residuals remain on the order of $10^{-1}$ p.u. in Figure~\ref{fig:quadratic_tightness}~\footnote{Residuals are reported in per unit on a common \SI{100}{\mega\watt} base.}. 

These observations motivate a final refinement step, in which an \ac{AC}-\ac{PF} is solved using the optimization output as initialization. Mean, median and maximum \ac{AC}-\ac{PF} solution times per time step (with reactive-limit checks) are respectively \SI{0.0906}{\second}, \SI{0.0892}{\second}, and \SI{0.1396}{\second} confirming that the refinement step remains computationally tractable across all considered horizons.
Finally, since in the decomposed scheme subproblems are solved independently of the main problem, tightness can be promoted without careful tuning of loss-penalty weights, while preserving an interpretable cost structure for storage investment. In contrast, centralized formulations often require disproportionately large weights on losses for them to dominate the objective~\cite{yi_optimal_2023}.
\vspace{-3mm}
\subsection{Optimal siting and sizing}
As a final simulation, the proposed framework is tested in a typical \ac{TSO} setting. The planning horizon spans $365$ days, the candidate \ac{BESS} locations are restricted to the $12$ buses exhibiting voltage magnitude violations and the ampacity limits are set to 82\% of the maximum branch current achieved in the results of the \ac{PF} for the branches with candidate battery locations at both ends ($10$ in total). The convergence threshold is set to $\varepsilon=5\mathrm{e}{-}3$, weights to $w_{\text{loss}} = 1$, and $w_{\text{slack}} 1\mathrm{e}{+}3$. Without loss of generality, costs are unitary $IC_s^{\text{power}} = IC_s^{\text{energy}} = 1$ CHF/\SI{}{\watt} and CHF/\SI{}{\watt\hour}, $\>\forall s$, $OC_l= 1$ CHF/\SI{}{\watt^2}, $\>\forall l$. Table~\ref{tab:final_results} reports the resulting optimal siting and sizing decisions together with the corresponding objective value. The algorithm terminates in 148 Benders iterations, for a total solution time of \SI{16145}{\second} (less than \SI{5}{\hour}). Infeasibility is manifested in $39$ subproblems and feasibility is finally restored after iteration $143$ out of $147$. The total time spent solving the power flow once convergence is achieved is \SI{974}{\second} and tightness is of the same order of magnitude of Figure~\ref{fig:quadratic_tightness}.
 \begin{table}[H]
    \centering
    \scriptsize
    \setlength{\tabcolsep}{5pt}
    \renewcommand{\arraystretch}{1.12}
    \caption{Total number of installed batteries, their rated power and energy, CAPEX and OPEX.}
    \label{tab:final_results}
    \resizebox{\linewidth}{!}{
    \begin{tabular}{lcccc}
        \toprule
        \textbf{$\sum_{s\in \mathscr{S}} U_s$} & \textbf{$\sum_{s\in \mathscr{S}} W_s$ } & \textbf{$\sum_{s\in \mathscr{S}} C_s$ }& \textbf{CAPEX } & \textbf{OPEX}\\
        \midrule
        $12$ & \SI{1576}{\mega\watt} & \SI{5488}{\mega\watt\hour} & $70.6$ CHF & $10980.3$ CHF \\
        \bottomrule
    \end{tabular}}
    
\end{table}
\vspace{-3mm}
\section{Conclusion}\label{sec:conclusion}
In conclusion, the proposed framework enables the computationally tractable solution of high-resolution, operationally driven \ac{BESS} allocation problems over annual (and, in principle, multi-year) horizons on large-scale meshed transmission networks. By reformulating the planning task as a \ac{MISOCP} and solving it via a parallelized \ac{GBD} scheme, solution times are reduced to the range of minutes to hours (depending on the boundary conditions) even on portable hardware. The formulation is general and can be specialized to different operational objectives (in this work, efficiency is targeted under voltage- and ampacity-limit infeasibilities), and features rigorous convergence guarantees. Moreover, an \ac{AC}-\ac{PF}-feasible operating point can be recovered from the relaxed solution at modest additional computational cost. Scalability results indicate that long horizons are primarily limited by available parallelism, whereas the dominant bottleneck is the number of linking variables, motivating a sparse candidate \ac{BESS} locations set, an assumption that is often consistent with practical planning tasks.

Overall, the framework supports \acp{TSO} considering the procurement of \ac{BESS} as grid-support assets: with market-cleared net injections as input, it enables congestion management and voltage regulation with \ac{TSO}-controlled resources that are constrained to be market-neutral, potentially reducing the need for costly remedial actions.

Future work will extend the assessment to larger, real-world (non-synthetic) transmission networks, include $N-1$ security assessment and broaden scalability studies, including scenario variability and stochastic formulations.

\bibliographystyle{IEEEtran}
\vspace{-3mm}
\bibliography{references.bib}
\vspace{-5mm}
\appendix\label{sec:appendix}
\subsection{Convergence for mixed-integer master problem variables}
Finite convergence of the \ac{GBD} algorithm is established here for a master problem comprising both integer and continuous variables, with the continuous variables restricted to a convex set. In~\cite{geoffrion_generalized_1972}, finite convergence is guaranteed if either (i) the master-feasible set $Y$ is finite, or (ii) $Y$ is a compact (i.e., closed and bounded) subset of the feasible set of the linking variables in all subproblems, denoted by $V$. In brief, case (i) follows from the finiteness of $Y$ and from the fact that the master problem cannot generate the same solution more than once. Case (ii) relies on the existence and uniform boundedness of the optimal dual multiplier vectors of the subproblems for all $y\in Y$, which is ensured if Slater’s constraint qualification holds at each such point. Consequently, if all values of the linking variables that lead to infeasible subproblems are eliminated in finitely many iterations, and if the remaining master-feasible set is a compact subset of $V$, finite convergence follows.

For the problem considered in this work, subproblem infeasibility can only occur when the master sets the siting decision to zero at a candidate location, i.e., when a \ac{BESS} is not installed. Since the number of such binary configurations is finite, all infeasible siting patterns can be excluded in finitely many iterations. After their exclusion, the remaining master-feasible region is a finite union of Cartesian products of bounded continuous domains (2D boxes for sizing variables) with binary decisions. Hence, it is contained in a bounded hyper-rectangle and is therefore a compact subset of $V$. This implies finite convergence of \ac{GBD} by Theorem~1 in~\cite{geoffrion_generalized_1972}.
\vspace{-3mm}
\subsection{Feasibility check subproblem formulation}
Consider the feasibility-check subproblem \text{(FSBP)} as formulated in~\cite{shahidehpour_benders_2005} (see the reference for the notation). Note that this construction extends to conic constraints via the generalized Farkas lemma, as shown in~\cite{VectorConicConstraint}.
\begin{equation}
\text{(FSBP)}:  
    \begin{aligned}
        & \min_{\mathbf{x}  \geq \mathbf{0}} &  \mathbf{1}^{\mathrm{T}} \mathbf{s} & \\
        & \text { s.t. } & \mathbf{E x} + \mathbf{Is} \geq \mathbf{h}-\mathbf{F} \mathbf{y} & \quad \quad :\mathbf{u}\\
        & & \mathbf{y}  = \hat{\mathbf{y}}  & \quad \quad :\mathbf{\rho}\\
        & & \mathbf{s} \geq \mathbf{0} & \quad \quad :\mathbf{\lambda} 
    \end{aligned}
\end{equation}
The authors argue that $\text{(FSBP)}^\prime$ is a particular instance of $\text{(FSBP)}$, which yields the same feasibility cuts formulation.
\begin{equation}
\text{(FSBP)}^\prime:  
    \begin{aligned}
        & \min_{\mathbf{x}  \geq \mathbf{0}} & \mathbf{1}^{\mathrm{T}} \mathbf{s} & \\
        & \text { s.t. } & \mathbf{E x} \geq \mathbf{h}-\mathbf{F} \mathbf{y} & \quad \quad :\mathbf{u}^\prime\\
        & & \mathbf{y}  = \hat{\mathbf{y}} + \mathbf{s} & \quad \quad :\mathbf{\rho}^\prime\\
        & & \mathbf{s} \geq \mathbf{0} & \quad \quad :\mathbf{\lambda}
    \end{aligned}
\end{equation}
Feasibility cuts follow from strong duality: for a minimization problem, the dual objective provides a lower bound on the primal optimum. Moreover, subproblem feasibility (equivalently, zero optimal slack) implies that the primal optimal value equals $0$. Therefore, subproblem feasibility requires $\inf_{\mathbf{x}\geq \mathbf{0},\,\mathbf{y},\,\mathbf{s}} \mathcal{L}(\mathbf{x},\mathbf{y},\mathbf{s}) \le 0$. The derivation of the left-hand side of this inequality constraint proceeds identically for both $\text{(FSBP)}$ and $\text{(FSBP)}^\prime$, with the only difference being a linear transformation of the slack variables by $\mathbf{F}$. The latter is less general, as it enforces $\dim(\mathbf{s})=\dim(\mathbf{y})$ rather than matching the dimension of the constraint residuals; consequently, infeasibilities that cannot be resolved by relaxing only the linking variables cannot be represented in formulation $\text{(FSBP)}^\prime$.

The Lagrangian function associated with $\text{(FSBP)}$ is given by \eqref{eq:fk}.
\begin{equation}\label{eq:fk}
    \begin{aligned}
        \mathcal{L}(\mathbf{x},\mathbf{y},\mathbf{s}) =& \mathbf{1}^{\mathrm{T}} \mathbf{s} + \mathbf{u}^{\mathrm{T}}\left(\mathbf{h}-\mathbf{F} \mathbf{y} - \mathbf{E x}-\mathbf{I} \mathbf{s} \right) \\
        & + \mathbf{\rho}^{\mathrm{T}} \left(  \mathbf{y}-\hat{\mathbf{y}}\right) - \mathbf{\lambda}^{\mathrm{T}}  \mathbf{s},
    \end{aligned}
\end{equation}
This yields the first-order optimality conditions \eqref{eq:KKT} and complementarity slackness conditions \eqref{eq:dual_opt}.
\begin{equation}\label{eq:KKT}
    \begin{aligned}
        & \nabla_{\mathbf{x}\geq 0} \mathcal{L}(\mathbf{x},\mathbf{y},\mathbf{s}) = - \mathbf{E}^{\mathrm{T}} \mathbf{u} = 0\\
        & \nabla_{\mathbf{y}} \mathcal{L}(\mathbf{x},\mathbf{y},\mathbf{s}) = \mathbf{\rho} - \mathbf{F}^{\mathrm{T}} \mathbf{u}=0\\
        & \nabla_{\mathbf{s}} \mathcal{L}(\mathbf{x},\mathbf{y},\mathbf{s}) = \mathbf{1} - \mathbf{I} \mathbf{u}- \mathbf{\lambda}=0,
    \end{aligned}
\end{equation}
\begin{equation}\label{eq:dual_opt}
    \begin{aligned}
        & \mathbf{h}-\mathbf{F} \mathbf{y} - \mathbf{E x}-\mathbf{I} \mathbf{s}=0, &  \mathbf{u}>0\\
        & \mathbf{s}>0, &  \mathbf{\lambda}=0.
    \end{aligned}
\end{equation}
The dual objective is then given by \eqref{eq:complementarity}, which yields the feasibility cut $\mathbf{\rho}^{\mathrm{T}} \left(\mathbf{y} - \hat{\mathbf{y}}\right) + v(\hat{\mathbf{y}})\leq 0$.
\begin{equation}\label{eq:complementarity}
    \begin{aligned}
        \inf_{\mathbf{x}\geq 0,\mathbf{y},\mathbf{s}} \mathcal{L}(\mathbf{x},\mathbf{y},\mathbf{s}) &= \mathbf{u}^{\mathrm{T}}\mathbf{h}-\mathbf{\rho}^{\mathrm{T}} \hat{\mathbf{y}} \\
         &=\mathbf{u}^{\mathrm{T}}\left(\mathbf{F} \mathbf{y} + \mathbf{E x}+\mathbf{I} \mathbf{s}\right)- \mathbf{u}^{\mathrm{T}}\mathbf{F} \hat{\mathbf{y}} \\
         &=\mathbf{u}^{\mathrm{T}} \mathbf{F} \left(\mathbf{y} - \hat{\mathbf{y}}\right) + \left( \mathbf{1}- \mathbf{\lambda}\right)^{\mathrm{T}} \mathbf{s}\\
         &=\mathbf{\rho}^{\mathrm{T}} \left(\mathbf{y} - \hat{\mathbf{y}}\right) + \mathbf{1}^{\mathrm{T}} \mathbf{s}.
    \end{aligned}
\end{equation}

% The same derivation applied to $\text{(FSBP)}^\prime$ yields the first-order opitmality condition with respect to the slack variable,
% \begin{equation}\label{eq:KKT}
%     \begin{aligned}
%         & \nabla_{\mathbf{s}} \mathcal{L}(\mathbf{x},\mathbf{y},\mathbf{s}) = \mathbf{1} - \mathbf{F} \mathbf{u}- \mathbf{\lambda}=0
%     \end{aligned}
% \end{equation}
% leading to the dual objective
% \begin{equation}\label{eq:complementarity}
%     \begin{aligned}
%         \inf_{\mathbf{x}\geq 0,\mathbf{y},\mathbf{s}} \mathcal{L}(\mathbf{x},\mathbf{y},\mathbf{s}) &= \mathbf{u}^{\mathrm{T}}\left(\mathbf{F} \mathbf{y} + \mathbf{E x}+\mathbf{F} \mathbf{s}\right)- \mathbf{u}^{\mathrm{T}}\mathbf{F} \hat{\mathbf{y}} \\
%          &=\mathbf{\rho}^{\mathrm{T}} \left(\mathbf{y} - \hat{\mathbf{y}}\right) + \mathbf{1}^{\mathrm{T}} \mathbf{s}
%     \end{aligned}
% \end{equation}

\end{document}